\RequirePackage{fix-cm}
\documentclass[smallextended]{svjour3}

\usepackage{graphicx}
\usepackage{amssymb}

\usepackage{lineno}
\usepackage{verbatim}						
\usepackage{url}

\usepackage[utf8x]{inputenc}
\usepackage[colorinlistoftodos]{todonotes}
\usepackage[colorlinks=true, allcolors=blue]{hyperref}

\begin{document}


\title{The implications of digitalization on business model change}

\author{Magnus Wilson  \and  Krzysztof Wnuk \and Lars Bengtsson}

\institute{Magnus Wilson \at
              Blekinge Institute of Technology, BTH, Karlskrona, Sweden \\
              Ericsson AB, Karlskrona, Sweden \\
              Tel.: +46 455 38 50 00\\
              \email{magnus.wilson@bth.se}           
           \and
           Krzysztof Wnuk \at
              Blekinge Institute of Technology, BTH, Karlskrona, Sweden \\
              Tel.: +46 455 38 50 00\\
              \email{krzysztof.wnuk@bth.se}           
            \and
            Lars Bengtsson \at
              Lund Univeristy, LTH, Lund, Sweden \\
              Tel.: +46 46 222 00 00             \\
              \email{lars.bengtsson@design.lth.se} 
            }

\date{Received: date / Accepted: date}



\maketitle
\begin{abstract}

Context: Digitalization brings new opportunities and also challenges to software companies. 

Objective: Software companies have mostly focused on the technical aspects of handing changes and mostly ignoring the business model changes and their implications on software organization and the architecture. In this paper, we synthesize implications of the digitalization based on an extensive literature survey and a longitudinal case study at Ericsson AB. 

Method: Using thematic analysis, we present six propositions to be used to facilitate the cross-disciplinary analysis of business model dynamics and the effectiveness and efficiency of the outcome of business modeling, by linking value, transaction, and organizational learning to business model change. 

Conclusions: Business model alignment is highlighted as a new business model research area for understanding the relationships between the dynamic nature of business models, organization design, and the value creation in the business model activities.

\end{abstract}



\section{Introduction}
Digitalization brings new opportunities manifested as new business models. Increased connectivity is the main fuel for digitalization and Ericsson is one of the leaders of the networked socitey transformation \footnote{https://www.ericsson.com/en/reports-and-papers/networked-society-insights}. The advent of the 5G network stands is a prominent example of opportunities and challenges associated with massive connectivity when all value-chain members and partners must rethink or reorganize their positions if necessary. For many companies, 5G will revolutionize their business models and often force them to redefine their business offerings. However, with this speed of technological change, the business models can not remain static or re-actively responding to changes. 

This paper discusses the implications of digitalization on the transactional nature of business models \cite{Legner2017} based on an extensive literature survey and a longitudinal case study at Ericsson.  We synthesize six propositions for improved handling of business model change and add solution-oriented details to facilitate a cross-disciplinary discussion of broader implications on business-model research \cite{Ritter2017}. 

Digitalization drives significant changes to the process level, organization level, and business level of any company and their customers, as well as on the society level \cite{Parviainen2017}. Digitalization offers a significantly shorter turnaround-time on a transaction. As a consequence, the increased transaction speed drives new challenges for the alignment of business and technology changes. 

Companies are undergoing significant transformations and are struggling with the alignment of business and technology changes \cite{Snihur2017,Curado2006,Legner2017,Tongur2014}. Until recently, companies handled increasing size and complexity by: 1) clearly distinguishing between the planning and realization layers for company strategy, product portfolios and individual products; and 2) handling change mainly in the realization layer and ensuring that the planning layer remains reasonably stable. 

Digitalization increases the speed of change in the planning layer, which in many cases, reaches the speed of changes in the realization layer. As a result, negotiation, and risk management can no longer only rely on the sales and engineering departments, as the business models shift focus to the ecosystem and collaboration \cite{Moore1998}, \cite{Zott2010a}, \cite{Romero2011} and companies choose operating multi-business-models \cite{Snihur2017}. Business modeling literature also recognize the need for efficiently handling change as several authors discuss the dynamic nature of business models and change in the business environment, e.g., \cite{Fjeldstad2018,Cosenz2018,Saebi2014,Cavalcante2014,Hoflinger2014a,Hacklin2012,Romero2011,McGrath2010a,Doz2010,Chesbrough2010,Ballon2007,Osterwalder20051}, just to name a few. 


This paper contributes to the business model body of knowledge in three ways. The first contribution is a detailed solution-oriented, cross-disciplinary synthesis on the digitalization's impact on the alignment between business and technology change. We propose an extension to Ritter \& Lettl's business model research framework \cite{Ritter2017} with \textit{Business Model Change (BMCh)} based on the \textit{Value Membrane (VaM)} and \textit{Learning Organization Design (LOD)}, to facilitate the analysis of business model dynamics and impacts on effectiveness and efficiency. As the second contribution, we synthesize six propositions aimed at setting the context for a fusion between the practices of business modeling (BM) and requirement engineering as a new evolving practice \textit{Digital Business Modeling (DBM)}\footnote{The term was first introduced by an SAP White paper, Digital Business Modeling: A Structural Approach Toward Digital Transformation, DOI: 10.13140/RG.2.2.22643.73766/1}. Thirdly, we provide a list of consequences for industry and a cross-disciplinary research agenda derived from our synthesis.

The paper is structured as follows: In Section 2, we present our synthesis based on background and related work. In section 3 we present how the business environment changes for our industry case and our findings from the longitudinal study. In section 4, we summarize and discuss our results using the derived value membrane concept and develop one additional proposition. In section 5, we conclude our paper.

\section{Background and Related work}

The synthesis provided in this section is based on an extensive systematic literature review about efficiency,  effectiveness \cite{Wilson2018a} 
and flexibility of business modeling \cite{Wilson2018b} 
published in our previous work. It is also derived from our design science study on how to capture changing \textit{business intents} using \textit{context frames} \cite{Silvander2017a}. 
Our synthesis responds to multiple requests for cross-disciplinary research agenda \cite{Ritter2017}, \cite{Zott2013}, \cite{Veit2014}, \cite{Fjeldstad2018}, just to name a few. With six propositions, we summarize vital concepts, such as business model and BM, value creation and value capture, VaM, LOD, BMCh, strategy and business plans, layered and tiered architectures, business flexibility (BF).


Building on Ritter \& Littl's business model research framework, we are inspired by their argumentation that the alignment perspective offers the significant contribution to the academic discourse and their analogy for the business model as a \textit{membrane} between theories \cite{Ritter2017}. By analyzing uncertainty and equivocality \cite{Eriksson2016} with (missing) value in a transaction, as the membrane between two actors in an activity system \cite{Zott2013}, we propose the business model can also act as the "contextual agent" in what we call the \textit{value membrane (VaM)}. The VaM can help to identify the cause of the misalignment, and facilitate minimizing gaps between needed change, planned change, and implemented change. We propose the VaM to be seen as a value surface between the context frame interactions of two actors \cite{Silvander2017a}. 

 Our literature review confirms that most scholars either focus on detecting or preparing change at one level (strategy, portfolio, or product), or analyzing the broader aspects of the organization, external environment, and innovation without integrating the activities \cite{Wilson2018a}. 
 Many scholars are calling for further research on change realization, e.g., \cite{Osterwalder20051,Ballon2007,Veit2014}. Meier and Bosslau argue that there is almost no attention in research to the dynamic aspects, flexibility, validation, and implementation of business models \cite{Meier2012a}, while Richter et al. emphasize the importance of understanding the degree of flexibility needed to realize change \cite{Richter2010}. Seeing business models as activity systems helps organizations (as responsible for the business) adapt to change and generate value \cite{Fjeldstad2018}. 
 



\section{Results}
We have synthesized five   propositions based on the literature review results and one based on the case study. We applied thematic synthesis of the previous literature review results \cite{Wilson2018a}, as it allowed for identifying and analyzing patterns (themes) within data. Thematic synthesis helps to interpret various aspects of the research topic \cite{6092576}. Each of the selected 57 papers was analyzed in detail and relevant chunks of text were marked and assigned to one of few categories: digitalization, value transformation, business model change, business flexibility, abstraction layers in business model change. Next, we constructed interpretations in each area and explored the relationships between the five themes (areas). Our high-order factors became the propositions presented in this paper. Finally, we evaluated the trustworthiness of the interpretations.  

\subsection{The impact of digital transformation on the nature of negotiating a business deal and equivocality}\label{sec:BizNeg}

With the digital transformation of the business environment \cite{Bharadwaj2013a}, \cite{Matt2015}, \cite{Legner2017}, negotiation, and risk management can no longer rely on the sales and engineering departments, but need to enact business model changes towards ecosystem and collaboration \cite{Moore1998}, \cite{Zott2010a}, \cite{Romero2011}. The negotiating power, coming from: (1) knowing what \textit{business flexibility (BF)} can be offered; (2) how BF is translated into a contractual flexibility that can be absorbed by the business model realization (partners, organizations, and business processes); (3) without jeopardizing the underlying effectiveness and efficiency of products and technical solutions (promised contractual characteristics); emerges as a critical competitive advantage. However, with more roles participating in the negotiation \cite[Figure 7 p.1182]{Silvander2017a},
uncertainty and equivocality (multiple and conflicting interpretations of a goal, situation, or task) can negatively impact  quality, cost, and lead-time of both the planning and realization phases \cite{Eriksson2016}, \cite{Koufteros2005}, \cite{Chang2006}.  

Companies undergoing the digitalization transformation should detect if the previously used realization strategy (the combination of the business model, products and services) still will adhere to the changed contractual terms and conditions. This involves checking if the current business model will accommodate the new terms and conditions, and the associated risks to deliver the changed contractual terms. The distance between strategizing, innovating, and planning for BMCh is significantly reduced.  We argue that such risk management should be done before signing any contract, and therefore propose that,\newline  

\textbf{Proposition 1: A mechanism for early-detection of business model change is a critical factor in maintaining a company's negotiating power to ensure business success, via improved risk management derived from the business flexibility.}

\subsection{The gap between Business Model Planning and Execution}\label{sec:BizModel}

Business model experimentation is gaining more importance for software companies, as a response to a growing need for business model innovation \cite{Chesbrough2010} and digitalization \cite{Legner2017}.  Experimentation is an approach to achieve effective change to the business, driven by the rationale that in ``\textit{highly uncertain environments, strategies are about insight, rapid experimentation, and evolutionary learning as much as the traditional skills of planning and rock-ribbed execution}" \cite{McGrath2010a}.


To analyze the gap between planning and execution, we complement Höfflinger's top-down definition of the business model with Rohrbeck et al. bottom-up definition of business modeling, ``\textit{to be a creative and inventive activity that involves experimenting with content, structure, and governance of transactions that are designed to create and capture value}" \cite{Rohrbeck2013}. 


Rohrbeck et al. focus on experimenting, as a 'round-trip' process of 'translating an idea into execution, test, evaluate, and change until satisfied' (similar to the agile method of developing software products followed up by proper retrospectives). Secondly, they also focus on transactions, thereby connecting the business model to human behavior and value in execution and planning activities. Thirdly, they make a clear distinction between created value and captured value, as two (role-dependent) views of a transaction, implying an information representation suitable for maintaining (observe, analyze, decide, change) many relationships  to support effective and efficient collaborations (through all the stages of the business model lifecycle, e.g., plan, design, deployment, execution, phase out).

Inspired by Fjeldstad \& Snow, we adopt the idea of value as the contingency variable affecting all other elements of the business model \cite{Fjeldstad2018}, and to understand the transaction- and role-dependent \textit{Direction of Value (DoV)}, we build on the value concept proposed in the Value Delivery Metamodel (VDML) \cite{VDML2015}. We also adopt the terminology introduced by the Software Value Map (SVM) \cite{Khurum2013a}.

 Osterwalder proposes BM to include the following capabilities: (1) Understand and share, i.e., Capture, Visualize, Understand, Communicate and share; (2) Analyze, i.e., Measure, Observe, Compare; (3) Manage, i.e., Design, Plan, Change \& Implement, React, Align, and Improve decision-making; and (4) Prospect, looking into the future, i.e., Innovate, Business model portfolio, Simulate and test \cite{Osterwalder2004b}. These four groups of capabilities, together with the four knowledge areas proposed by Fjeldstad \& Snow provide us the boundaries of the practice of BM.

However, neither H\"offlinger \cite{Hoflinger2014a}, Fjeldstad \& Snow \cite{Fjeldstad2018}, nor VDML \cite{VDML2015} makes a clear separation between value creation and value capture. Also, neither Osterwalder \cite{Osterwalder2004b,Osterwalder2015} nor Zott et al \cite{Zott2010a} make an unequivocal distinction on what level value is discussed in the value creation and value capture processes. We, therefore, propose that \newline

\textbf{Proposition 2: Value translation and value transformation capabilities are essential for BM. By exploring value, in an interactions on the individual level as the unit of analysis, we can resolve ambiguities in relation to the different areas of the business model (e.g., product offering, product delivery, product development, finance, customer relationships, partner management) stemming from: (1) the \textit{direction of value}; (2) inter-level relationships of source and target for value; and (3) aggregation issues for value creation and value capture (scalability and value slippage). }

\subsection{Handling Business Model change}\label{sec:BMCh}

Both radical or incremental business model changes need  \cite{Laasch2018} to be addressed both at the planning and the realization levels \cite{Cavalcante2014}. Cavalcante et al. \cite{Cavalcante2011} divided BMCh into four types of change: business model creation; extension; revision; and termination. They further argued there is a 'pre-stage' of 'potential of BMCh' before the actual change occurs, often including analysis, experimentation, and other activities to build insights, learning, and commitment. In software engineering, this phase would include extensive prototyping or building the minimum viable product. Therefore, he proposes to develop a detailed guide for analyzing BMCh, both at the level of cognition as well as action, where he sees continuous experimentation and learning as fundamental pillars for effective BMCh, transforming the company into a `\textit{permanent learning laboratory}`.

To address change on the planning level, a company needs to understand the As-Is situation, (which capabilities exist), and the effects on the To-Be situation (needed abilities) due to an 
identified misalignment. Such insights require understanding how strategy relates to business model, tactics, and residual choices  \cite{Casadesus-Masanell2010a}, in combination with what strategic agility \cite{Doz2010} and level of strategic flexibility \cite{Schneider2014} the organization has.

To facilitate such insights, we propose to represent a business model, by combining the work by Ghezzi's on value networks (VN) and resource management (RM) \cite{Ghezzi2013}, with Osterwalder's business model canvas (BMC) \cite{Osterwalder2010}. Therefore, a company's need for BMCh can be derived from having profound knowledge and a sound understanding of the three dimensions: (1) the customer(s) and related relationships; (2) the value proposition (revenue streams, what values to create, how to deliver it to the customer); and (3) the company's assets (products, resources, activities, cost structures, and partner relationships). 


To address change on the realization level, i.e., solutions implemented in products, processes, and organizations, literature discuss concepts like business model operationalization (BMO), implying reconfiguration and tuning of the company's assets \cite{Ballon2007}, aligning business with IT \cite{Salgado2014a,Salgado2014b}, business model experimentation \cite{Chesbrough2010}, \cite{McGrath2010a}, collaborative business modeling \cite{Romero2011}, Dynamic Software Product lines \cite{Capilla2014}, R\&D as innovation experiment systems \cite{Holmstrom2013}, just to name a few. With the advent of the digital business strategy \cite{Bharadwaj2013a}, we propose that,\newline

\textbf{Proposition 3: Software companies possess a unique advantage for detecting and implementing BMCh. Using their software development process to integrate their business model innovation with their product innovation, they can efficiently develop 'native' product support for managing the linkage of contractual flexibility to the configuration of software products, to achieve richer levels of business model experimentation and collaborative business modeling.}

\subsection{Increasing Business Flexibility}

Flexibility helps organizations to  ``\textit{adapt when confronted with new circumstances...and provides the organization with the ability to respond quickly to market forces and uncertainty in the environment.}`` \cite{Lucas1994}.  Richter et al. points out that embedding flexibility into system design can address risks in relationships and optimize stakeholder's incentives, turning incomplete contracts into opportunities \cite{Richter2010}. They discuss \textit{changeability} as a term to better understand investments in flexibility related to value, cost, and risk. Changeability is defined by options under internal ('robustness' and 'adaptability') respectively external control ('flexibility' and 'agility'). 

In the business and management literature, flexibility is discussed in many different contexts, as related to business models and as ways to managed change, e.g., strategic flexibility \cite{Mason2012a,Schneider2014}, resource and organizational flexibility versus dynamic capabilities \cite{Barney1991}, \cite{Sanchez1995}, \cite{Teece1997}, \cite{Ghezzi2012}, and business model flexibility\cite{Mason2012a,Mason2008e}. 


We define Business Flexibility (BF), as the ``\textit{negotiable options in: 1) Relationship; 2) Financial; and 3) the Value proposition between two parties trying to reach an agreement}". These options are needed for an effective negotiation to leverage a company's ability to compromise without breaking the promise in the final contractual agreement. The terms Relationship, Financial, and Value proposition refer to the context of Osterwalder's right side of the BMC \cite{Osterwalder2010}. Using the BMC, a company visualizes the strategic decisions and critical business options that characterizes the rationale of the business idea, and how it strategy-wise will be turned into a successful business (model) realization. 

A change (on planning or realization level) is  triggered by a gap (misalignment) in expectations and what is delivered. Closing these gaps (transforming a capability into an efficient ability) requires significant investments in time and effort, involving a multitude of collaborating roles (internal and external). Also, closing the gap adds an extra dimension to the notion of flexibility, as for how to realize a solution \cite{Capilla2014}, \cite{Richter2010}. 

\textbf{Proposition 4: Software companies have a unique opportunity for implementing business flexibility and efficiently creating value propositions. 
Software companies should develop software architectures and software functionality to enable a synchronized change in their business model.}

\subsection{The 3-layer BMCh Abstraction model}\label{sec:3LayerModel}
 Casadesus-Masanell \& Ricart argued a clear distinction between strategy and the business model, where the business model ``\textit{is a reflection of the firm's realized strategy}" and that the strategy is the plan and process to reach the desired goal via the business model and onto tactics \cite{Casadesus-Masanell2010a}. Strategy refers to the choice of the business model while Tactics refer to the residual choices open to the company. 

The Business Motivation Model (BMM) \cite{BMM_OMG2015} separates the business plan into three layers, ENDS, MEANS, and Realization. Using BMM plus a business model, a company could model their total operations, from company vision, via several business models, business processes, and down to individual tasks and rules guiding each decision and choice. Such schemas are in use and also heavily researched, e.g.; TOGAF\footnote{The Enterprise Architecture standard used by the world’s leading organizations to improve business efficiency, http://www.opengroup.org/subjectareas/enterprise/togaf}, MEMO \cite{Frank2014}. 

Separating the top layer descriptions from the bottom layer with a middle layer, provides separation of concerns and increases the maintainability of each layer by limiting impacts due to changes in the other layers. Such pattern can be found in many different contexts, 
e.g. Ends, Goals, Means chains \cite{Gutman1997}; Presentation layer, Business layer, Data layer in Software architectures\footnote{see Microsoft Application Architecture Guide, 2nd Edition, Chapter 1, 3, and 5, http://msdn.microsoft.com/en-us/library/dd673617.aspx}
; Business Item Library, Capability Library, Practice Library \cite{VDML2015}; Business architecture, IS architecture, and Technology architecture\footnote{Three layers of architecture to support the requirement management in accordance with the architecture vision. The Open Group Architecture Framework (TOGAF), http://www.opengroup.org/subjectareas/enterprise/togaf}.

However, when combining such patterns into a conceptual model, contextual ambiguity becomes a challenge since each layer may discuss a 'topical' view, rather than a strictly defined Tier that can be distinctly separated. Contextual ambiguity can result in layers overlapping, creating new dependencies, resulting in unforeseen consequences and gaps in the contextual model. 


To minimize contextual ambiguity, we build on BMM, TOGAF, MEMO, and VDML to define our 3-layer BMCh Abstraction model (BMCh AM) as \textit{'Business layer', 'Capability layer', and 'Realization layer'}. We use TOGAF's definition of Capability ``\textit{as an ability that an organization, person, or system possesses. Capabilities are typically expressed in general and high-level terms and typically require a combination of organization, people, processes, and technology to achieve.}``. 

By using BMCh AM, we consequently only use a Capability to describe an Ability. A Capability should not include any realization, allowing for different options to perform the ability, e.g., outsourced, as tasks in activities and processes, automated business processes, by machines, humans or mixed, e.g. the Amazon Mechanical Turk\footnote{See https://www.mturk.com/}. We, therefore, by combining BMM, TOGAF, and VDML, with 'context frames' \cite{Silvander2017a}, %
propose that \newline


\textbf{Proposition 5: Given our 3-layer BMCh AM \textit{'Business layer', 'Capability layer', and 'Realization layer'}, we can conceptualize BMCh as '\textit{a gap between BF, efficiency, and value}'.}

\section{Case study: adapting to the digital transformation in the telecommunication industry}\label{sec:Ericsson}

For Ericsson AB\footnote{https://www.ericsson.com/en}, one critical aspect of achieving the business and technology transformation and managing change, has been a long-term focus on industrialization and automation of the product development and the delivery (via process innovation). Ericsson has shown remarkable adaptability and flexibility during the past 50 years of technology growth and software (engineering) evolution. Two critical strategies laid the foundation for this success. By (1) actively driving the telecommunication standards, the business environment could be kept reasonably stable, enabling (2) an efficient industrialization built on the core processes of developing, testing, and delivering software-intensive products and solutions. 

However, digitalization requires additional strategies for handling the fast-paced business environment than driving technology standards. The technology innovation must be in concert with an equally dramatic and accelerating business model innovation. Ericsson's business model has evolved from the resource-centric, standard product-sales model, via several product and service models, over into different use models, where software-intensive products and services now are sold and delivered as-a-service and on demand. Today, Ericsson is running multi-business-model operations, and with that, facing additional challenges to keep up with the pace of change. A majority of these challenges can be structured according to Ritter \& Lettl's framework,  aiding the understanding of risks related to effectiveness, efficiency, and misalignment due to temporal effects related to uncertainty and equivocality.  

\subsection{Business model change at Ericsson}
Digitalization shifted the business risks to new dimensions, e.g. business ecosystem (sharing and collaborating in fierce competition), rather than optimizing the own company's assets as a part of a value-delivery chain (e.g., traditionally mitigating risks with long-term business agreements and international standards). Such BMCh, profoundly impacts the financial steering and control, as much of the investments need to be taken up-front, while the majority of revenues shifts to on-demand usage rather than sales of products \cite{Meier2012a,Richter2010}. More importantly, the transition from business models based on selling products or hourly-rated services (with a strong focus on add-on sales), into value-based, knowledge-intensive, customer-unique use-models, has affected many of Ericsson's dynamic and strategic capabilities and most of the core business processes. 

For Ericsson, this also impacted the organizational design, requiring extended focus on organizational learning and incentives, governance and management structures suited for the inherent dynamics, as well as collaborating with strategic and operational information. It also required enhanced clarity in responsibility and authority for the business model activities.

\subsection{The longitudinal study (2012-2016), a global program for industrializing services}

Back in 2012, the Ericssons' service organization, established in 2007-2008, was mainly working in two types of business models: 
\begin{itemize}
\item Managed Services - running the operator's network for them with large, long-term contracts.
\item Service consultancy and Delivery model - focused on project deliveries and learning services.
\end{itemize}

As part of a corporate strategy realization to put the customer first, the service organization devised their strategic program "Global Scale - Local Reach", involving 75000+ resources (global, regional, and contractors) in nine regions, working in three segments of the service portfolio (Managed Services, Product Related Services, and Consulting and System Integration). The goal of the program was to improve customer responsiveness, improve productivity, and improve internal benchmarking.  

We conducted a longitudinal case study between 2012-2016, where we actively worked alongside teams responsible for 
\begin{itemize}
\item supporting the program manager and his steering group with a business and enterprise architecture analysis, 
\item responsible for the business level requirements towards tools and IT development, and 
\item consultants for the deployment (business processes and training) into the sales and delivery organization (global plus nine regions). 
\end{itemize}

In the beginning of the program (2012-2013), we participated in eleven extensive workshops interviewing practitioners from affected areas: finance; product management (services and software products); key account managers; Ericsson IT (master data, business processes, and system responsibles); sales; delivery (project); and support processes (planning, development, and pricing, of services). The 3-4 hours workshops were based on a short introduction to the workshop and the program, followed by practitioners presenting their current business processes and ways of working. Practitioners were then interviewed on current issues and potential opportunities was discussed under the frame of the new program, providing us with great insights of the scope plus the strategical and the operational issues facing the program. The workshops also provided a deeper understanding of the level of uncertainty, equivocality, and rivalry between the different roles and organizations. We were also given continuous access to all program-related information, monthly reports, and steering group protocols. 

To identify any misalignment against the program's (original) goals and the actual outcome in the deployment and to understand the longitudinal effects of the program (2012 and 2016), we also conducted two sets of individual, 60+ minutes interviews, with a delivery project manager and a solution architect. 
 
As a pilot, Ericsson applied the industrializing of the sales and delivery processes in 30+ deliveries to customers in three regions during 2013. These pilot projects delivered contract scoping efficiency and accuracy improvement by 88\% - first time right. The ordering process was considered simplified, while delivery lead time, and project costs were reduced by 12-35\%. However, the program complexity and program duration were significantly underestimated (duration  exceeded by 150\%). We identified three main reasons for the increased complexity: 
\begin{itemize}
\item the scalability of the piloted solution turned out a bigger issue than anticipated. 
\item the inherent complexity (flexibility and re-usability) of the services to be industrialized and the services' dependency on the skills and knowledge of the service delivery staff.
\item frequent re-organizations - this could be traced back to a substantial BMCh together with an insufficient support for fast and cross-organizational learning, negatively impacting the transformation program. 
\end{itemize}


The program struggled with two major challenges: 1) to decide what services to industrialize and which should remain 'customer-specific' (due to required customer variability vs. investing in standard product options vs. a too high dependency on the skills and knowledge of the service delivery staff), 2) to find the best balance for the new and updated IT tools to minimize disruptions to operations while concurrently updating the business processes.   

The technical solution to the first challenge was basically divided in five parts, with a need for completely new tools to be integrated with existing tools and processes. 
The second challenge 
proved to be complex mainly due to the volume of tacit and explicit information in various forms of knowledge representations, and realizations with efficient knowledge management systems. 

Related to the second challenge, the decisions between investing in tool support versus investing in business process flexibility (requiring more skilled staff and investments in more options in the products) turned out to be very challenging, mainly due to the multi-disciplinary value argumentation presentation to the decision makers and top management. As a consequence, the 'traditional' IT update and integration process of new and existing tools to match the evolving business processes, was affected by misunderstandings and delays leading to temporary solutions in the sales and delivery organization. Under customer pressure to deliver on signed contracts, this led to decreased trust between organizations, affecting the efficiency of the collaboration.

It also proved difficult to synchronize the business process development (sales and delivery processes to use industrialized services) with the agile Ericsson product development (the new generation of products to be delivered using the updated business processes). We identified the following four root causes of the misalignment:
\begin{itemize}
\item temporal effects due to different life cycles of these two core business processes, 
\item organizational steering, coordination and incentives, 
\item expected capabilities that did not deliver on the requested abilities in customer projects, and 
\item the differences between the old and new product generations, the needed training of the service delivery staff, and their valuable customer experience feedback to the R\&D organization.
\end{itemize}


\subsubsection{Temporal effects of organizational learning}

The temporal effects of organizational learning created a gap between feed-forward and feedback loops. The different organizations (R\&D, sales, delivery, Ericsson IT) were occupied with their life-cycles of change as committed in earlier plans, see \cite[Figure 5]{Silvander2017a}. %
The symptoms of this were observed in areas of communication, coordination, training, and reporting, resulting in uncertainty, equivocality, and sub-optimization at best and a lack of abilities at worst. 

Scaling the solution was affected since planned capabilities needed by different organizations were not translated (in time) into required abilities, i.e., integrated tools and staff adequately trained in relation to the new or changed business processes (so they could perform the tasks demanded by the evolving business model). The scale of the industrialization problem was among the most significant factors since it affected the amount of information and the relationships between the affected organizations involved in the change processes. The rippling change-reaction escalated and started to violate existing goals, commitment, and reporting, leading to more efforts spent on temporary, local solutions to assure customer contracts could be honored.

\subsection{Case Study Results Summary and Synthesis}


Ericsson's traditional, engineering-centered industrialization approach, would have benefited by categorizing the strategic program's requirements and associated risks into the five areas (strategic decisions, business logic, business model artifacts, misalignment, and BMa)
and highlighting that the program was actually facing a BMCh. By addressing the misalignment between the effectiveness ('do the right thing' as a top-down strategic planning process) and the efficiency (as the bottom-up change of existing BMa, business processes, organizations, and tools), we believe the scale of the program, as well as the temporal affects, could have been predicted and managed in a better way by proposing a set of different tactics (stemming from a BMCh), thereby invoking a higher degree of top management commitment and attention.  

The majority of the issues are connected to the effectiveness area, and in particular related to misjudging the temporal affects, when reaching a common understanding (minimizing equivocality) of the goals and tactics to accommodate the new goals with existing organizational goals. This study confirms opportunities and challenges for  digitalization reported by scholars, for example, \cite{Meier2010a,Chew2014,Richter2010,Snihur2017}. 

The case study also highlights the added complexity of BMCh for large software companies that operate with contracts spawning years to complete. This calls for a combination of BMCh and organizational design. What appears to be inevitable is that the business environment will change during the execution of the underlying agreements. Our interview respondents believed that governance mechanisms should facilitate the exploration phase (Knowledge Creation process), transforming tacit knowledge into explicit knowledge fast enough and made it available through the Knowledge Management process.

We believe it requires fast, efficient Feed-forward and Feedback loops between R\&D, sales, and the service delivery organizations, illustrating the continuous interaction between Knowledge Creation and Knowledge Management processes.  Support for these loops should preferably be implemented both in the products as well as in the business processes. We therefore propose that,\newline

\textbf{Proposition 6: The practice of Digital Business Modeling (DBM) should be coined as a fusion between current practices of business modeling and requirement engineering, and become a key practice in LOD to facilitate business model innovation through experimentation.}\newline

\section{Conclusion}

Many distinguished scholars have highlighted the cross-disciplinary complexity stemming from the on-going digitalization and transformation of the business environment \cite{Veit2014,Romero2016,Legner2017,Zott2013} to name a few. This paper highlights three critical aspects of business modeling in the analysis of the misalignment between planning and execution. Firstly, focus on experimenting \cite{Rohrbeck2013}, as a 'round-trip' process of 'translating an idea into execution, test, evaluate, and change until satisfied' (similar to the agile method of developing software products followed up by proper retrospectives). Secondly, focus on transactions \cite{Zott2010a}, thereby connecting the business model to human behavior and value in execution and planning activities. Thirdly, the analysis is direction-sensitive, with minimum two (role-dependent) views of the transaction, implying an information representation suitable for maintaining (observe, analyze, decide, change) many relationships (supporting efficient collaborations through all the stages of the business model lifecycle, e.g., plan, design, deployment, execution, phase out) \cite{Silvander2017a}. 

We propose Business Model Change (BMCh) as an extension to Ritter \& Lettl’s business model research framework and its' business model alignment\cite{Ritter2017}. We believe that addressing BMCh by linking it to Learning Organization Design via the Value Membrane could address many of the identified cross-disciplinary challenges, as indicated by Legner et al. ``\textit{digital transformation requires a focus on the business solution first.., the foundations for the technological system background should be laid, rather than vice versa}`` \cite{Legner2017}. 

This paper is an initial step for such a detailed, cross-disciplinary guide for handing BMCh. Synthesizing from two literature reviews ,  \cite{Wilson2018a,Wilson2018b} 
a design science study \cite{Silvander2017a}, %
and the case study presented in this paper, we present six propositions for addressing the challenges of aligning the planning and execution layers for software-intensive product development (software). We also highlight four critical aspects that software companies need to address: 

\begin{itemize}
	\item Business model innovation for the business ecosystem, e.g., driven by markets and contextual changes, co-creation of value, collaboration within and between organizations, partners, communities, and customers, new streams of revenue while sharing of risks, revenues, and costs \cite{Romero2011,Rohrbeck2013}.
	\item A digital software, focused on automation and integration of business and software architecture, information, and tools. Products designed for mass-customization in a collaborative agile development, may sustain the speed of change and increasing demand of customer experiences (delivered as cloud services) \cite{Olausson2010,Capilla2013,Magnusson2017}.
	\item Badly designed organizations, ill-suited for experimentation and collaboration in a digital business world, affecting both the product development as well as the value delivery, e.g., agreement structures, incentives, processes, knowledge management and organizational learning, measurements of effectiveness and efficiency, revenues, cost, decision-making based on multifaceted optimization and transparency \cite{ReInventOrg2014,Kahneman2011}.
    \item The level of integration and automation between the four processes of value creation, value capture, knowledge creation, and knowledge management \cite{Lepak2007,Curado2006}. This is the foundation for an innovative enterprise and should be nurtured as a key competitive advantage.   
\end{itemize}

\section*{ACKNOWLEDGMENT}
We would like to acknowledge that this work was supported by the KKS foundation through the S.E.R.T.  Research Profile project at Blekinge Institute of Technology. 




\section*{REFERENCES}
\bibliographystyle{IEEEtran} 
\bibliography{main.bib}






\end{document}